\documentstyle[12pt]{article}
\begin{document}

\newcommand{\be}{\begin{equation}}
\newcommand{\ee}{\end{equation}}
\newcommand{\ba}{\begin{eqnarray}}
\newcommand{\ea}{\end{eqnarray}}
\newcommand{\dalam}{\raisebox{1mm}{\fbox{}{}}\;}
\newcommand{\pa}{\partial}
\newcommand{\f}{\frac}
\newcommand{\st}{\stackrel}
\newcommand{\s}{\sqrt}


\author{A.A.~Logunov ~and~ M.A.~Mestvirishvili\\
\small{Institute for High Energy Physics, Protvino, 142 281, Russia}}
\title {The Causality Principle \\in the Field Theory of Gravitation}
\date{}
\maketitle

\begin{abstract}
The causality principle for the Relativistic Theory of
Gravitation (RTG) is presented. It is a straightforward
consequence of the RTG basic postulates. The necessary  conditions for
physical solutions of the gravitational field equations  to be fulfilled are
given.
\end{abstract}


The Relativistic Theory of Gravitation~[1] (RTG) as the field theory of
gravitation is based on a hypothesis, that the gravitational field, as well
as all other fields, propagates in the Minkowski space, and its source is such a
universal conserving quantity as the energy-momentum tensor of all the matter,
including the gravitational field itself. This approach allows one to uniquely construct a theory of the gravitational field as a gauge theory (within the framework of
second order equations).

A complete set of the RTG gravitational equations in a system of units 
$
\hbar=c=G=1 $ 
looks like
\be
\gamma^{\alpha\beta}D_\alpha D_\beta\tilde\Phi^{\mu\nu}
+m^2 \tilde\Phi^{\mu\nu}=16\pi t^{\mu\nu}, \ee
\be
D_\nu\tilde\Phi^{\mu\nu}=0.
\ee

Here
$ \gamma ^ {\alpha\beta} $ 
is a metric tensor of the Minkowski space in
arbitrary coordinates;
$ \tilde\Phi ^ {\mu\nu} = \s {-\gamma} \Phi ^ {\mu\nu} $ is a density of the
gravitational field tensor;
$D_\mu $ is the covariant derivative of the Minkowski space;
$m $ is a rest mass of the gravitational field;
$t^{\mu\nu}$ is a density of the energy-momentum tensor of all matter.

Density of the matter energy-momentum tensor $t ^ {\mu\nu} $ 
consists of a
density of the gravitational field energy-momentum tensor  $t ^ {\mu\nu} _g $
and a density of the substance energy-momentum tensor  $t_M ^ {\mu\nu} $. We
call the substance all the matter fields, except for the gravitational field
\be
t^{\mu\nu}=t^{\mu\nu}_g+ t^{\mu\nu}_M.
\ee
The interaction of a gravitational field and a substance is taken into account
in the energy-momentum tensor  density of  substance $t ^ {\mu\nu} _M $. The
gravitational field differs from all the known fields in  that the gravitational
interaction affects the terms with higher (second) derivatives, whereas all
other fields do not enter terms with second derivatives. This feature of
the gravitational field also gives rise to the effective Riemannian space.
Density of the energy-momentum tensor $t ^ {\mu\nu} $, according to  Hilbert,
can be expressed through the Lagrangian  scalar density  $L $ as follows:
\be
t^{\mu\nu}=-2\f{\delta L}{\delta\gamma_{\mu\nu}}, \ee
where
\be
\f{\delta L}{\delta\gamma_{\mu\nu}}= \f{\partial L}{\partial\gamma_{\mu\nu}}
-\partial_\sigma
\left ( \f{\partial L}{\pa \gamma_{\mu\nu,\sigma}} \right ),\;\;
\gamma_{\mu\nu,\sigma}=\f{\pa\gamma_{\mu\nu}}{\pa x^\sigma}.
\ee
Eqs. (1) and (2) can be derived from the principle of least action
only if the Lagrangian  density is  taken as follows~[1]:
\be
L=L_g+L_M(\tilde g^{\mu\nu},\;\Phi_A).
\ee
Here
\be
\tilde g^{\mu\nu} = \tilde \gamma^{\mu\nu}
+\tilde \Phi^{\mu\nu}, \ee
$$
\tilde g^{\mu\nu}=\s{-g} g^{\mu\nu}, \;\; \tilde
\gamma^{\mu\nu}=\s{-\gamma}\gamma^{\mu\nu},\;\;
\Phi_A \;\;\;\mbox{--- fields of substance}.
$$
The  Lagrangian density of the gravitational field  $L_g $ in (6) is  ~ [1]
\be
L_g=\f{1}{16\pi}
\tilde g^{\mu\nu}
(G^\lambda_{\mu\nu}
G^\sigma_{\lambda\sigma}
-G^\lambda_{\mu\sigma}
G^\sigma_{\nu\lambda})
-\f{m^2}{2}
\left (
\f{1}{2}\gamma_{\mu\nu}
\tilde g^{\mu\nu} -\s{-g} -\s{-\gamma} \right ).
\ee
Here $G ^\lambda _ {\mu\nu} $ is a third rank  tensor 
\be
G^\lambda_{\mu\nu}=\Gamma^\lambda_{\mu\nu}
-\gamma^\lambda_{\mu\nu}
=\f{1}{2}
g^{\lambda\sigma}
(D_\mu g_{\sigma\nu}
+D_\nu g_{\sigma\mu}
-D_\sigma g_{\mu\nu}).
\ee
$ \Gamma ^\lambda _ {\mu\nu} $ are the Christoffel symbols of the effective Riemannian space,
$ \gamma ^\lambda _ {\mu\nu} $ are the Christoffel symbols of the  Minkowski space.

The densities of Lagrangians (6) and (8) are the results of initial standings of
the theory. On the basis of (6) it is evident, that the motion of substance in the
Minkowski space under impact of a gravitational field is reduced to a
motion of substance in the effective Riemannian space. Thus, the effective
Riemannian space appears as a straightforward consequence of the hypothesis,
that a source of a gravitational field is the conserving energy-momentum tensor
of matter .

As the gravitational field (for example) in an inertial system is determined in
one frame, then according to Eq. (7)  the effective Riemannian space  is also
determined in one frame, but it means, that it has a prime topology. Thus, the
effective Riemannian space arising due to the impact of the gravitational field
in the  Minkowski space always has a prime topology. In the General Theory of
Relativity (GRT) the complicated topologies of Riemannian spaces are possible,
and therefore atlas of maps is necessary for its exposition. The field notions
about gravitation are rather strong, and therefore they do not allow one to derive
the GRT  equations on their basis.

From Eqs. (6) and (8) and from the principle of least action we see that  Eqs. (1)
and (2) can be written as follows~[1]:
\be
R_{\mu\nu} -\f{m^2}{2}
(g_{\mu\nu}-\gamma_{\mu\nu})
=8\pi
\left (
T_{\mu\nu}-\f{1}{2}g_{\mu\nu}T\right ), \ee
\be
D_\nu\tilde g^{\mu\nu}=0.
\ee
Here
\be
T^{\mu\nu}=-2\f{\delta L_M}{\delta g_{\mu\nu}},\;\;
T_{\mu\nu}=g_{\mu\sigma}g_{\nu\lambda}T^{\sigma\lambda}.
\ee

Eqs. (1) and (2), as well as Eqs. (10) and (11), are covariant under arbitrary
coordinate transformations and are form-invariant concerning Lorentz
transformations. The last statement means, that they precisely obey the Special
Principle of  Relativity. According to the RTG, the inertial forces and  the forces of
gravitation are separated, they are of different nature. If the inertial forces
can be removed by a choice of the  inertial system of coordinates, the forces of
gravitation cannot be eliminated by a choice of a frame even locally. J. Synge 
wrote on this subject almost half century ago~[2]:  ``{\it In  Einstein's
theory the gravitational field is present or absent depending on the Riemannian 
tensor being nonzero or  equal to zero. This property is absolute; it is  in no
 way related to a world line of an observer.}''

Considering the
gravitational field as a physical field in the Minkowski space, we should with
necessity observe the Causality Principle in the Minkowski space. Its essence
is in the following: For a moving test body in the Minkowski space it is always
possible to pick such a frame, in which this body will be at rest, whatever is 
the nature of forces this movement  would be called by,  and consequently,
the requirement 
\be d\sigma^2 =\gamma _ {00} (x) (dx^0) ^2 > 0, \ \; \gamma _{00} (x) > 0. \ee 
should take place. In Minkowski's paper ``Space and time'' ~[3], published in 1909, this standing was formulated as follows:  ``{\it We
shall introduce now the following postulate. \underline{A substance present at any world
point can always be viewed as being at rest under a}\\ \underline{suitable definition of space and time}}''.

But taking into account that in the Minkowski space the test body goes along a
geodesic line of the effective Riemannian space by impact of a gravitational
field, the following requirement should also be fulfilled
\be
ds^2=g_{\mu\nu}(x)dx^\mu dx^\nu>0, \ee
that in a frame where the body is at rest becomes:
\be
ds^2=g_{00}(x)(dx^0)^2>0,\;\;
g_{00}(x)>0.
\ee
Thus, for a motion of a test body in the Minkowski space it is always  possible
to pick such a frame, in which this body is at rest, but simultaneously the
requirements of  causality (13) and (15) should be fulfilled.
\ba
ds^2=g_{00}(x)(dx^0)^2>0,\nonumber \\ \\ [-0.3cm]
d\sigma^2=\gamma_{00}(x)(dx^0)^2>0. \nonumber \ea
These causality  requirements can be written in the following form~[1]:
\be
\gamma_{\mu\nu} (x)U^\mu U^\nu=0, \ee
\be
g_{\mu\nu} (x)U^\mu U^\nu\leq0. \ee
In writing these requirements we also take into account the opportunity of
presence in  nature of particles with a zero rest mass, which motion happens
along isotropic geodesic lines. Requirements (16), as well as Eqs. (17) and
(18) mean that the causality cone of the effective Riemannian space should be
positioned inside the causality cone of a Minkowski space. If the causality cone
of the Riemannian space was beyond the causality cone of the Minkowski space,
that  resulted  in the impossibility of picking in the Minkowski space a frame, in
which the test body  be able to stay at rest. It would mean that a
three-dimensional force of gravitation of such   a ``gravitational field''  is
impossible to balance by any inertial force, as in this case the following
inequality  would take place:
\be
d\sigma^2=\gamma_{\mu\nu} (x)dx^\mu dx^\nu<0.
\ee
From here it follows, that such a ``gravitational field'' cannot be presented as a physical field propagating in the Minkowski space. By virtue of the causality 
requirements  Eqs. (17) and (18) the effective Riemannian space will possess an
isotropic and timelike geodesic completeness. 

The Causality Principle also  ensures the
existence of a spacelike surface in  the Riemannian space, which is only
once  intersected by any non-spacelike curve, i.e. there is a global Cauchy
surface, to set initial physical conditions for this or that problem. In
solving the Hilbert-Einstein equations one selects only such solutions, for
which the following requirement takes place at any point of space: 
\be
g<0,
\ee
and also for any timelike vector $K_\nu $ the following inequality is fulfilled:
\be
T^{\mu\nu}K_\mu K_\nu\geq 0, \ee
and the quantity $T ^ {\mu\nu} K_\nu $ for a given vector $K_\nu $ should  form
a non-spacelike vector.

In the RTG in solving Eqs. (10) and (11) it is necessary to select only such
solutions, which alongside with requirements (20) and (21) also obey the
causality  requirements (17) and (18). The causality  requirements do not
follow from the equations, but this is not unusual, as also in  electrodynamics
the Causality Principle is not a corollary of Maxwell-Lorentz equations, it is
introduced in the form of Eq.~(13) as a supplement.  From Eqs.~(10) and (11) it
follows, that the test body moves along a geodesic line of the effective
Riemannian space. This line is defined by
\be
\f{dp^\nu}{ds} + \Gamma^\nu_{\alpha\beta} p^\alpha p^\beta=0,\;; p^\nu =
\f{dx^\nu}{ds},\;\;
ds^2=g_{\mu\nu}dx^\mu dx^\nu>0.
\ee
Here Christoffel symbol $ \Gamma ^\nu _ {\alpha\beta} $ is 
$$
\Gamma^\nu_{\alpha\beta}=\f{1}{2}g^{\nu\sigma}
(\pa_\alpha g_{\sigma\beta}+\pa_\beta g_{\sigma\alpha} -\pa_\sigma
g_{\alpha\beta}).
$$
According to the RTG, such motion is not free, as it occurs in  the Minkowski space
under impact of a gravitational field force.  A concept of gravitational force
is  absent in the GRT. J. Synge  wrote so related to this subject~[2]:
 ``{\it In the Theory of Relativity the concept of force of gravitation is absent,
as the gravitational properties are naturally build in the structure of
space-time and are exhibited in a space-time curvature, i.e. in that the
Riemannian  tensor $R _ {\mu\nu\lambda\sigma} $ is different from zero}''.

In the RTG the concept of force of gravitation is preserved, as the gravitation is
obliged to the existence of a gravitational field in the Minkowski space. Below we
shall determine this force, basing on Causality Principle (17) and (18) and
following  Ref.~[4]. According to the definition of the covariant  derivative
in the Minkowski space, we have
\be
\f{Dp^\nu}{ds}=
\f{dp^\nu}{ds}
+\gamma^\nu_{\alpha\beta}p^\alpha p^\beta.
\ee
By using (22) in (23), we discover
\be
\f{Dp^\nu}{ds}=-G^\nu_{\alpha\beta}p^\alpha p^\beta.
\ee
Let us present the l.h.s. of Eq. (24) in the following form:
\be
\f{Dp^\nu}{ds}
=\left (
\f{d\sigma}{ds}
\right )^2 \left [
\f{DV^\nu}{d\sigma}
+V^\nu
\f{\f{d^2 \sigma}{ds^2}} {\left (
\f{d\sigma}{ds}
\right )^2} \right ],\;\; V^\nu =\f{dx^\nu}{d\sigma}.
\ee
Here $V ^\nu $ is a timelike four-vector of velocity in the Minkowski space. It
obeys  the following condition:
\be
\gamma_{\mu\nu}V^\mu V^\nu=1,\;\;
d\sigma^2>0.
\ee
Substituting (25) in (24), we shall receive
\be
\f{DV^\nu}{d\sigma}
=-G^\nu_{\alpha\beta} V^\alpha V^\beta
-V^\nu
\f{\f{d^2 \sigma}{ds^2}} {\left (
\f{d\sigma}{ds}
\right )^2}.
\ee
On the basis of Eq. (26) we have
\be
\left (
\f{d\sigma}{ds}
\right )^2
=\gamma_{\alpha\beta} p^\alpha p^\beta.
\ee
By differentiating this expression over $ds $, we obtain
\be
\f{\f{d^2 \sigma}{ds^2}} {\left (
\f{d\sigma}{ds}
\right )^2}=
-\gamma_{\lambda\mu} G^\mu_{\alpha\beta} V^\lambda V^\alpha V^\beta.
\ee
By substituting this expression in Eq.  (27), we shall discover~[4]
\be
\f{DV^\nu}{d\sigma}=
-G^\mu_{\alpha\beta}
V^\alpha V^\beta (\delta^\nu_\mu -V^\nu V_\mu).
\ee
From here it is obvious, that the motion of a test body in the  Minkowski space
happens under impact of a four-vector of force $F ^\nu $
\be
F^\nu=
-G^\mu_{\alpha\beta}
V^\alpha V^\beta (\delta^\nu_\mu -V^\nu V_\mu), \;\;
V_\mu=\gamma_{\mu\sigma} V^\sigma.
\ee
It is easy to get convinced that
\be
F^\nu V_\nu =0.
\ee
The l.h.s. of  Eq. (30) by definition is equal to
\be
\f{DV^\nu}{d\sigma}
=\f{dV^\nu}{d\sigma}
+\gamma^\nu_{\alpha\beta} V^\alpha V^\beta.
\ee

It should be noted especially  that the motion of a test body along a
geodesic line of the effective Riemannian space can be understood as a motion
in the Minkowski space under impact of force $F ^\nu $, only if simultaneously the
following requirements take place
\ba
ds^2=g_{\mu\nu}dx^\mu dx^\nu >0,\nonumber \\[-0.3cm] \\
d\sigma^2=\gamma_{\mu\nu}dx^\mu dx^\nu >0,\nonumber \ea
i.e. if the Causality Principle is fulfilled.

The force of gravitation and the Riemannian curvature tensor are mutually
related. So, if the Riemannian curvature tensor is equal to zero, then by
virtue of Eqs. (10) and (11) the force of gravitation $F ^\nu $  will also be
equal to zero.  In case the curvature tensor  is different from zero,
then the force of gravitation  is not equal to zero as well. And on the contrary,
if the force of gravitation $F ^\nu $ is different from zero, then the
Riemannian curvature is not equal to zero either. The vanishing of force of
gravitation $F ^\nu $ results in equality to zero of the Riemannian curvature
tensor. Due to the impact of a gravitational  force $F ^\nu $ there occurs also  a
fall of a body in a gravitational field, i.e. everything is the same as  in
Newtonian physics. 

Moreover, all gravitational effects in Solar system
(deflection of a light beam by the Sun, time retardation of a radiosignal,
precession of a gyroscope, Mercury perihelion shift) are caused by impact of the
force of gravitation $F ^\nu $, instead of the Riemannian curvature tensor,
which in Solar system is small enough. It is explained by the fact that the
force of gravitation is determined by the first derivatives of the metric, whereas
the Riemannian curvature tensor~--- by the second derivatives. As the force of
gravitation $F ^\nu $ is a four-vector, it can never be converted to zero by a
choice of the frame.  Only the   three-dimensional part of the force of
gravitation $ \vec F $ can be converted to zero by a choice of the frame. It
also means that the gravitational field as a physical reality cannot be removed
even locally. 

The similar assertion takes place also for any physical field. A
system falling in a gravitational field is not even locally  inertial because
the motion of a test body along a geodesic line of the effective Riemannian
space is not free. If the test body was charged, it would radiate
electromagnetic waves, as it goes with acceleration. For this reason the
acceleration has an absolute meaning in the RTG, as the concept of inertial
systems of coordinates is maintained there.

 Authors of article [5] argue that the Causality Principle (17) and (18)  is
broken for a weak monochromatic wave in the linear approximation. However,
this conclusion is incorrect. Below we shall show, that everything is all right
with realization of requirements of  causality. Eqs. (10) and (11) in the
linear approximation of a perturbation theory take the following form:
\be
\gamma^{\alpha\beta} \pa_\alpha\pa_\beta
\Phi^{\mu\nu} + m^2\Phi^{\mu\nu}=8\pi T^{\mu\nu}, \ee
\be
\pa_\nu\Phi^{\mu\nu}=0.
\ee
Here $ \gamma ^ {\alpha\beta} = (1, -1, -1, -1) $. In the linear approximation
neither the interaction of a gravitational field with substance, nor the
self-interaction of a gravitational field are taken into account. The metric
tensor of  Minkowski space is   at higher derivatives in  Eqs. (35),
and consequently, the requirement of  causality for a gravitational field has a
standard form
\be
d\sigma^2=\gamma_{\mu\nu}dx^\mu dx^\nu>0.
\ee

It is especially noteworthy to mention that the system of equations (35), (36)
is physically inconsistent, as according to Eqs. (35) and (36) the 
energy-momentum  conservation law for substance takes place, on the one hand,
\be
\pa_\nu T^{\mu\nu}=0, \ee
and, on the other hand, there is a radiation of a gravitational field $ \Phi ^
{\mu\nu} $, which with necessity causes losses of substance energy and that
contradicts the energy-momentum tensor conservation law for the  substance  Eq.
(38). The effective Riemannian space, which follows from  Eqs. (35) and
(36), has the following metric
\be
g^{\mu\nu}=
\gamma^{\mu\nu}
+\Phi^{\mu\nu}
-\f{1}{2}\gamma^{\mu\nu}\Phi,\;\;
\Phi=\Phi^{\mu\nu}\gamma_{\mu\nu},
\ee
\be
g_{\mu\nu}=
\gamma_{\mu\nu}
-\Phi_{\mu\nu}
+\f{1}{2}\gamma_{\mu\nu}\Phi,\;\;
-g=1+\Phi.
\ee
Here $ | \Phi _ {\mu\nu} |, | \Phi | $  are small values in comparison with unity.
The occurrence of the effective metric of Riemannian space (40) results in the
test body (or graviton)  moving along a geodesic line of the effective Riemannian
space. It follows from the equation of this geodesic line that an integral of
motion arises
\be
g_{\mu\nu} p^\mu p^\nu=1,\;\;
p^\nu=\f{dx^\nu}{ds},\;\;
ds^2>0,
\ee
which takes into account an impact of the gravitational field on a test body
(or graviton). In Eqs. (35) and (36) these interactions are not taken into
account, and, therefore, there is a mutual coherence between the motion of a test
body (or graviton) and the definition of the effective Riemannian metric with  Eqs. (35) and (36). On the basis of Eq. (36) we get the following
relation for a monochromatic wave 
\be
p^\mu\Phi_{\mu\nu}=0.
\ee
Substituting (40) in (41) and taking into account (42), we get
\be
\gamma_{\mu\nu}\f{dx^\mu}{ds}\cdot
\f{dx^\nu}{ds}=
\f{1}{1+\f{1}{2}\Phi}>0.
\ee

As for any weak field the value of $ | \Phi | $ is small in comparison with unity
by virtue of  perturbation theory, it follows from here, that the timelike
vector $p ^\nu $ in the effective Riemannian space remains timelike also in the
Minkowski space, and consequently, the causality cone of  the effective
Riemannian space  is embedded inside the causality cone of the Minkowski space.
Thus, the Causality Principle (17) and (18) is fulfilled also for a weak monochromatic wave.

The introduction of Causality Principle in the form of Eqs. (16) or (17), (18)
is not an arbitrary requirement, but with necessity follows as a direct
corollary of a hypothesis, that the gravitational field, as well as all other
physical fields, propagates in the Minkowski space. Such an idea is implemented in
the RTG, and the metric tensor of the Minkowski space is contained in the initial
system of equations (10) and (11).

There is an assertion in the literature, that it is possible to present the GRT in
the form of a field theory by  using the Minkowski space. This is not true. Field
notions, as we have already mentioned, with necessity lead to the simple
topology of the effective Riemannian space, and also require the Causality
Principle in the Minkowski space to be fulfilled. But  all these
things are not found in the GRT and this is  its characteristic feature. Usage of the
Minkowski space  metric in the GRT is deprived of any  physical sense and
contradicts the logic of this theory. 

Field notions on gravitation together with
a hypothesis, that the conserving energy-momentum tensor is a source of the
gravitational field, with necessity lead us to the physical conclusion  of a
nonzero rest mass of the gravitational field, which was reflected in the
system of RTG equations. Thus, the presence of a nonzero rest mass of a
gravitational field follows from the  general enough  standings of the
theory~[1]: The gravitational field is a physical field propagating in the
Minkowski space similar to other physical fields, and a source of this field
is the universal conserving quantity --- the energy-momentum tensor of all the
matter including the gravitational field itself.

As there is a general similarity in construction between  Maxwell--Lorentz
electrodynamics and the RTG, it is natural to assume an opportunity of
existence of a nonzero rest mass also for a photon, as it has been mentioned
earlier~[6].
\vspace*{2mm}

The authors express their gratitude to V.I.~Denisov and Yu.V.~Chugreev for
valuable discussions.


\newpage
\begin{thebibliography}{9}
\bibitem{1}
 Logunov A.A. and Mestvirishvili M.A. The Relativistic Theory of Gravity. --- Moscow, ``Nauka'', 1989.
 \\ Logunov A.A. Relativistic Theory of Gravity. --- Nova Science
Publishers, Inc., vol.~215, 1998. Commack, New York.\\ Logunov A.A. Teoriya gravitacionnogo polya. --- Moscow, ``Nauka'', 2000.
\bibitem{2}
Synge J. General Relativity. --- Moscow, Izdatelstvo Inostrannoj Literatury, 1963.
\bibitem{3}
Minkowski H. Relativity Principle. --- Moscow, Atomizdat, 1973, p.~171.
\bibitem{4}
Rosen N. // Phys.Rev., vol.~57, 1940, pp.~147-153.
\bibitem{5}
Brian Pitts J. and Schieve W.C. gr-qc/0101058, v.~2, 2001.
\bibitem{6}
Logunov A.A. // Vestnik MGU, Ser.~3, Fizika, Astronomia, 1993, vol.~34, no.~4, pp.~3-19.

\end{thebibliography}
\end{document}